\documentclass[prb,preprint,groupedaddress,showpacs]{revtex4-1}
\usepackage{amsmath}
\usepackage{graphicx}
\usepackage{url}
\usepackage{hyperref}
\begin{document}
\bibliographystyle{apsrev4-1}
\title{Note on the numerical solution of the scalar Helmholtz equation in a nanotorus with uniform Dirichlet boundary conditions}
\author{N.D. Nguyen}
\email[E-mail me at: ]{NgocDuy.Nguyen@ulg.ac.be}
\author{R. Evrard}
\affiliation{D\'{e}partement de Physique B5, Universit\'{e} de Li\`{e}ge, B-4000 Li\`{e}ge, Belgium}
\author{Michael A. Stroscio}
\affiliation{Department of Electrical and Computer Engineering, University of Illinois at Chicago, Chicago, IL 60607 , USA}
\begin{abstract}
This note describes the solution of the Helmholtz equation inside a nanotorus with uniform Dirichlet boundary conditions. The eigenfunction symmetry is discussed and the lower-order eigenvalues and eigenfunctions are shown. The similarity with the case of a long cylinder and with that of the vibrations of a circular elastic membrane is discussed. This similarity is used to propose a classification scheme of the eigenfunctions based on three indices.
\end{abstract}
\pacs{02.60.Lj, 02.70.Dh, 02.70.Hm, 03.65.Ge, 73.22.Dj, 78.67.Bf, 78.67.Ch}
\maketitle
\tableofcontents
%
%
%******************************************************************************
%
\section{Introduction and symmetry considerations}
\label{int}
Over the past years, toroidal systems have drawn much interest in solid state physics, surface physics, and photonics in relation with the development of high-Q optical resonators \cite{amani}, optical nanoantenna \cite{tep}, and carbon nanotori \cite{liu}. The solutions of the Helmholtz equation in nanotori can be of great help in understanding the physical properties of photons, electrons, and phonons in systems with toroidal symmetry. For instance, the scalar Helmholtz equation gives the quantum states of electrons in semiconductor nanotori within the band-mass approximation and describes the states of confined polar phonons in the same systems. Despite its importance, a discussion of the methods of solving the scalar Helmholtz equation in toroidal systems, as well as a description of its solutions, cannot easily be found in the scientific literature. A first aim of the present note is to remedy this situation. Also, the authors of the present note have devoted a complete article \cite{des2} to the study of free-electron states and polar phonons in tori. A second aim of the present note is to give more details on the numerical computations involved in this latter work, Ref.\ \onlinecite{des2}. A third goal is pedagogical. The solutions of the Helmholtz equation in tori can be used as examples or illustrations in courses of lectures in mathematical physics or on numerical methods of solving partial differential equations.

We represent the torus in a system of axes with the origin at the torus center and the $Oz$ axis coinciding with the torus axis. The torus axial symmetry motivates writing the Helmholtz equation in cylindrical coordinates $r, z, \varphi$. The symmetry also entails that the solution of the three-dimensional Helmholtz equation can be written as
\begin{equation}
\psi_{n,\nu,m} (r, z, \varphi) = \phi_{n,\nu,m}(r, z) e^{i m \varphi}, \qquad m = 0, \, \pm 1, \, \pm 2, \, \pm 3, \, \ldots,
\label{h1}
\end {equation}
leading to the following two-dimensional equation
\begin{equation}
\frac {\partial^2 \phi_{n,\nu,m}(r, z)} {\partial r^2} + \frac {1} {r} \frac {\partial \phi_{n,\nu,m}(r, z)} {\partial r} + \frac {\partial^2 \phi_{n,\nu,m}(r, z)} {\partial z^2} + \left[ k_{n,\nu,m}^2 - \frac {m^2} {r^2} \right] \phi_{n,\nu,m}(r, z) = 0.
\label{h2}
\end {equation}
The boundary conditions determine what are the allowed eigenvalues $k_{n,\nu,m}^2$ and eigenfunctions $\phi_{n,\nu,m}(r, z)$ of this equation. In this note, we restrict ourselves to the case of uniform Dirichlet boundary conditions, which impose that the eigenfunctions vanish on the toroid surface. The meaning of the subscripts $n$ and $\nu$ is discussed in more detail in Sec.~\ref{comp}. As for $m$, its relation with the magnetic quantum number in atomic physics is obvious. As in atomic or molecular physics, the states corresponding to the different values of $m$ could be called $s$, $p$, $d$, \dots \ states in accordance with their symmetry; in fact, as the symmetry of the torus is not spherical, but cylindrical only, the notation $\sigma$, $\pi$, $\delta$, \dots \ would probably be more appropriate. In the case of tori, there is nothing equivalent to the azimuthal quantum number $l$.

The solutions of the Helmholtz equation depend on a single geometrical parameter $\alpha = r_M / r_m$, the ratio of the torus major radius to its minor one. We use the torus minor radius $r_m$ as unit of length, so that all the quantities in the Helmholtz equation are dimensionless. Accordingly, the unit of $k_{n,\nu,m}$ is $1 / r_m = 1$. Due to the existence of a symmetry plane perpendicular to the torus axis, the eigenfunctions of Eq.\ (\ref{h2}) are either symmetric or antisymmetric with respect to a reflection onto this plane. By analogy with molecular physics, we label the symmetric states with the superscript $+$ and the antisymmetric ones with the superscript $-$. It remains to decide on a further classification to differentiate the states with the same value of $m$ and the same reflection symmetry. This question of classification of the solutions of the Helmholtz equation in toroidal systems does not seem to have been previously considered. As this aspect is more easily discussed for particular examples, we first specify which methods we use to solve the equation and describe the solutions obtained with these methods.
%
%******************************************************************************
%
\section{Solving the scalar Helmholtz equation inside a torus}
\label{sol}
\begin{figure}[t]
\includegraphics{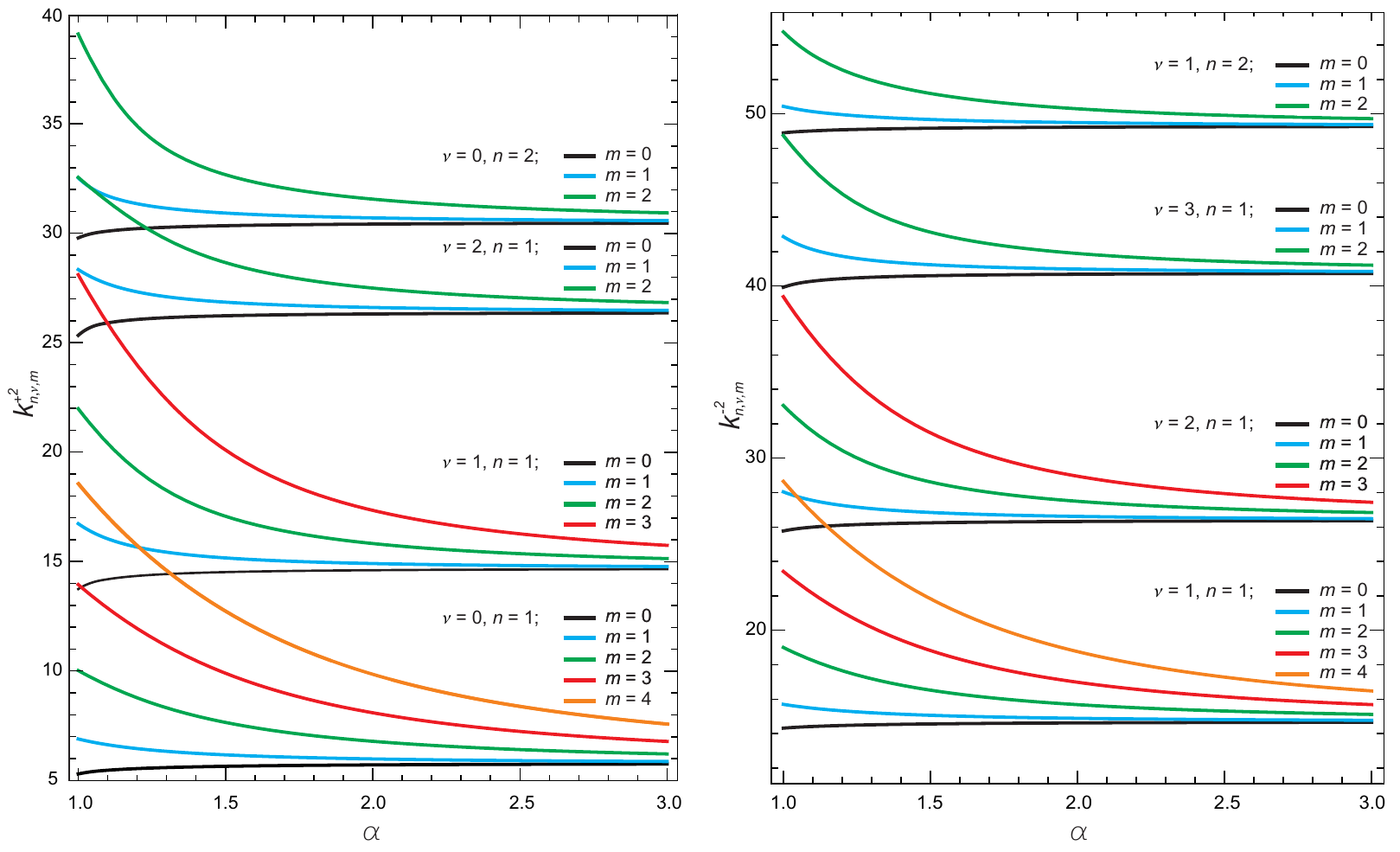}
\renewcommand{\baselinestretch}{1}
\caption{Lower-order eigenvalues of Helmholtz equation, in units of $1/r_m^2$, versus the ratio of the torus major radius to its minor one, $\alpha = r_M / r_m$. Left, eigenvalues $k^{+2}_{n,\nu,m}$ of symmetric solutions; right, eigenvalues $k^{-2}_{n,\nu,m}$ of antisymmetric solutions.}
\label{eigvl2}
\end{figure}
\normalsize
Solving the Helmholtz equation consists in finding its eigenvalues $k^2_{n,\nu,m}$ and the corresponding eigenfunctions $\phi_{n,\nu,m}(r, z)$. The Helmholtz equation does not separate in toroidal coordinates. This prevents obtaining analytic solutions in a torus. Therefore, we must resort to numerical methods of solving partial differential equations. We use either a spectral method \cite{spect} implemented in MATLAB or the finite-element method of MATLAB pdetool \cite{mlab}. Of course, both methods give the same results. The computer programs and many results are posted on the web page of Ref.\ \onlinecite{hlmwp}. Figure ~\ref{eigvl2} shows the fifteen low-order eigenvalues $k^{+ 2}_{n,\nu,m}$ for symmetric solutions and $ k^{-2}_{n,\nu,m}$ for antisymmetric ones, versus the ratio $\alpha = r_M / r_m$. The reason for labeling the solutions with two indices $n$ and $\nu$, besides $m$, is discussed in the next section. The allowed values of $n$ and $\nu$ are $n = 1$, 2, 3, \dots \ and $\nu = 0$, 1, 2, \dots, with the exception that the value $\nu = 0$ is not allowed in the case of antisymmetric solutions. Notice that the eigenvalues with the same $n$ and $\nu$, but different values of $m$, tend toward identical limits for $\alpha \rightarrow \infty$. These limits are the eigenvalues of the Helmholtz equation in a circular right cylinder of infinite length. This is discussed in more detail in the next section, Sec.~\ref{comp}. Also notice the crossing of some curves representing the eigenvalues as functions of $\alpha$. For example, in the case of symmetric modes, the curve giving the eigenvalue with $n = 1$, $\nu = 0$, and $m = 4$ intersects that of $n = 1$, $\nu = 1$, and $m = 0$ or $m = 1$ at values of $\alpha$ in the range from 1.2 to 1.3, approximately. This could be the source of interesting applications. For instance, in tori with appropriate dimensions, phonon-assisted transitions between two electron states with close energies could be used to populate states not attainable by direct optical transitions. In Ref.\ \onlinecite{des2}, we discuss in more detail the possible role of this near degeneracy in nanotori.
\begin{figure}[t]
\includegraphics[trim={2,3cm 7,5cm 2,3cm 7,7cm}]{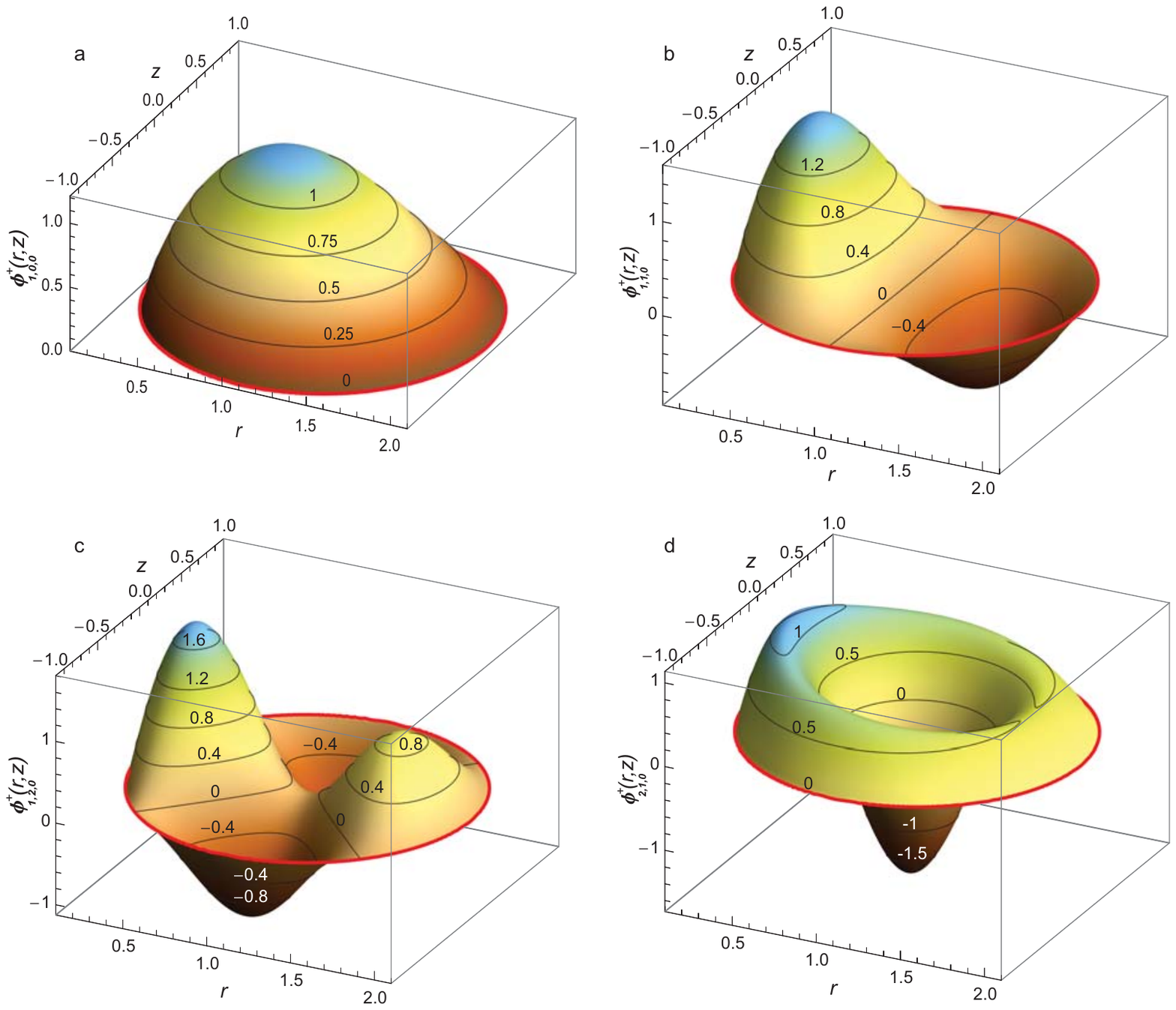}
\renewcommand{\baselinestretch}{1}
\caption{Low-order symmetric eigenfunctions of the Helmholtz equation in a torus with $\alpha = r_M / r_m = 1.1$, $m = 0$, and, respectively, a) $n = 1$, $\nu = 0$, b) $n = 1$, $\nu = 1$, c) $n = 1$, $\nu = 2$, and d) $n = 2$, $\nu = 0$, versus the position $r,z$ in the circle section of the toroid by the axial plane $\varphi = 0$. The eigenfunctions are squared normalized to 1. The black curves on the surfaces are contour lines labeled with the corresponding eigenfunction values. Of course, the contour lines $\phi = 0$ are nodal lines separating antinodes. The red curves show the boundary of the circular torus section.}
\label{sym4low}
\end{figure}
\normalsize

Figure~\ref{sym4low} shows graphs of the first four symmetrical solutions with the lowest eigenvalues, as functions of the cylindrical coordinates $r$ and $z$ in the circular section of the toroid by the axial plane $\varphi = 0$. These solutions correspond to $\alpha = r_M / r_m = 1.1$ and $m = 0$. Their squares are normalized to unity on the circular section. Contour lines are represented on the graphs, labeled with the corresponding values of the eigenfunction. In all the graphs of the eigenfunctions drawn in the present note, the red curve indicates the circular torus section. Recall that the torus axis of symmetry coincides with the $Oz$ axis. Many other solutions are shown on the web page of Ref.\ \onlinecite{hlmwp}.

The solutions with $\nu = 0$ show more or less concentric antinodes with a central one next to the circle center. The distance of this central antinode to the section center depends on the values of $m$ and $\alpha$. In general, the larger $m$ or the smaller $\alpha$, the larger the distance between the antinode and the section center. In the case of $\nu = 0$, the index $n$ gives the total number of antinodes. This is exemplified by the top left graph with $n = 1$, $\nu = 0$ and the bottom right graph with $n = 2$, $\nu = 0$, in Fig.~\ref{sym4low}. In the eigenfunctions with $\nu = 0$, there is no nodal line intersecting the circular boundary, so that these eigenfunctions keep the same sign in the whole region close to the circular boundary. This is in agreement with the fact that the value $\nu = 0$ is not allowed for antisymmetric solutions.

In the solutions with $\nu \neq 0$, the antinodes are located on more or less circular rings. The index $n$ gives the number of rings and $\nu$ gives the number of antinodes with the same phase, say the number of maxima, on each of these rings. As an example, the solutions in the graphs b and c of Fig.~\ref{sym4low} have a single antinodal ring implying that $n = 1$. The presence of a single maximum in the solution of graph b means that in this case $\nu = 1$ while $\nu = 2$ in the case of the solution in graph c.

All that precedes shows that $n$ plays a role similar to that of the principal quantum number in atomic physics while $\nu$ has some similarity with the magnetic quantum number. However, in the case of $\nu$, the similarity is only superficial since the center of the torus circular section is not a center of symmetry. For tori, the actual magnetic quantum number is the parameter $m$ previously defined.
\begin{figure}[t]
\includegraphics[trim={2.12cm 12.2cm 1.2cm 12.2cm},scale=0.93]{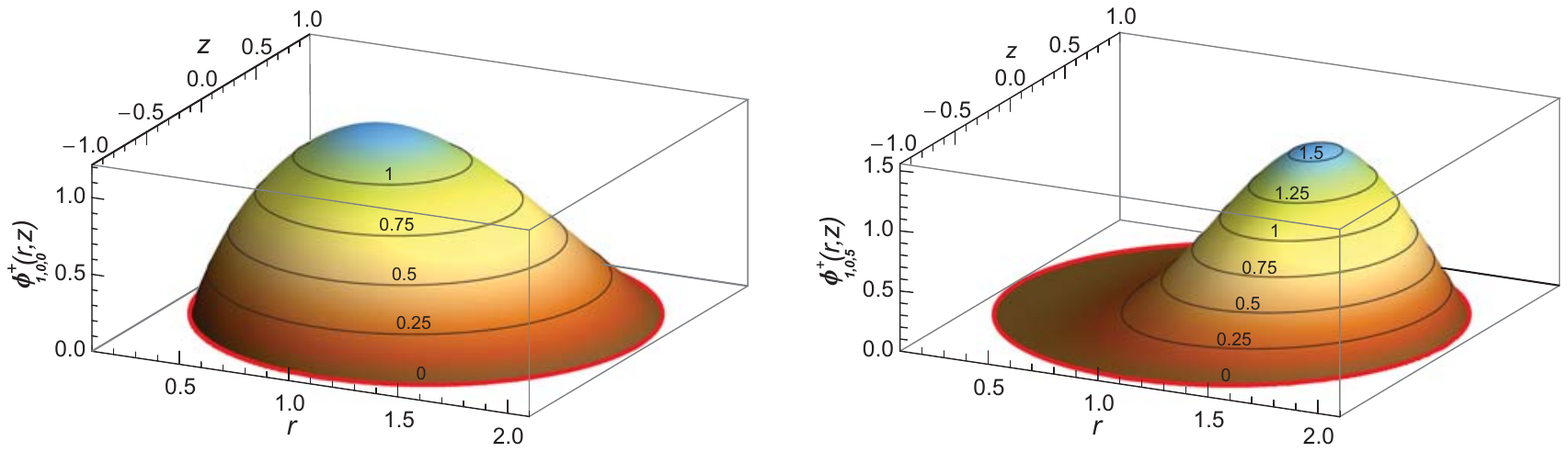}
\renewcommand{\baselinestretch}{1}
\caption{Comparison of two symmetrical eigenfunctions of the Helmholtz equation in tori with $n =1$, $\nu = 0$, and different values of $m$. Here again, $\alpha = 1.1$. Left, $m = 0$; right $m = 5$. The displacement of the eigenfunction maximum away from the torus axis with increasing values of $|m|$ is clearly visible.}
\label{difm}
\end{figure}
\normalsize

The shape of the eigenfunctions in the section of the toroid by the $\varphi = 0$ axial plane depends noticeably on the value of $m$. The position of the antinodes moves away from the torus axis with increasing absolute values of $m$, i.e., with increasing angular momentum. This is shown in Fig.~\ref{difm} which compares the eigenfunctions with $n =1$, $\nu = 0$, and either $m = 0$ or $m = 5$. Again, the value chosen for the ratio of the torus major axis to its minor one is $\alpha = 1.1$. In classical mechanics of massive particles or solids, the displacement away from the axis of rotation is due to the centrifugal force. As illustrated here, a similar effect exists in wave mechanics. This correspondence between classical particle mechanics and wave mechanics is a conceptual fact worth to be noticed.

The geometrical properties of the eigenfunctions of the Helmholtz equation in circular tori discussed here are in direct relation with that in the case of very long circular cylinders. Indeed, at the limit of a very large major radius to minor radius ratio, i.e., $\alpha \rightarrow \infty$, tori tend toward this type of cylinders. Therefore, the comparison with the case of cylinders should help to understand the geometrical features of the eigenfunctions in tori and the meaning of the $n$ and $\nu$ labels introduced here. This justifies that we now turn to a description of the solutions of the Helmholtz equation in long cylinders and to the comparison of these solutions with those in the case of tori.
%
%
%*****************************************************************************
\section{Comparison with the solutions in a circular cylinder}
\label{comp}
For $\alpha \rightarrow \infty$, the torus tends toward a right circular cylinder of length $l = 2 \pi r_M$. The torus major circumference becomes the cylinder axis of revolution and the section of the toroid by an axial plane tends to the section of the cylinder by a plane normal to the cylinder axis. This leads to write the Helmholtz equation in a different set of cylindrical coordinates $\rho, \zeta, \theta$ such that $O \zeta$ coincides with the cylinder axis and $\rho, \theta$ are the polar coordinates in the cylinder circular section. Contrary to the case of tori, the Helmholtz equation separates for cylinders. This is a classic problem in mathematical physics; see, e.g., Ref.\ \onlinecite{tang}, Sec. 6.3.4. For the sake of similarity with the case of tori, we impose periodic boundary conditions along the cylinder axis. Then, the solution is
\begin{equation}
\psi_c = e^{2i \pi m \zeta /l} \eta_{n, \nu} (\rho, \theta)
\label{c1}
\end {equation}
and the Helmholtz equation reduces to the following two-dimensional equation
\begin{equation}
\frac {\partial^2 \eta_{n, \nu}(\rho, \theta)} {\partial \rho^2} + \frac {1} {\rho} \frac {\partial \eta_{n, \nu}(\rho, \theta)} {\partial \rho} + \frac {1} {\rho^2} \frac {\partial^2 \eta_{n, \nu}(\rho, \theta)} {\partial \theta^2} + \left[ \kappa_{n, \nu}^2 - \left( \frac {2 \pi m} {l} \right)^2 \right] \eta_{n, \nu}(\rho, \theta) = 0.
\label{c2}
\end {equation}
The meaning of the indices $n$ and $\nu$ will become clear hereafter. For our purpose, we restrict ourselves to the case $l \rightarrow \infty$ with $m$ remaining finite. Then, the last term between the square brackets in Eq.\ (\ref{c2}) is negligible and the problem reduces to finding the time-independent part of the normal modes of vibration of an elastic circular membrane, which is a well-known problem in mathematical physics. For a time-dependent visualization of these modes, see Ref.\ \onlinecite{wikosc}. For more detail on the separation of variables in cylindrical coordinates, see the case of Laplace equation, Ref.\ \onlinecite{jack}, Sec. 3.7. The eigenvalues of Eq.\ (\ref{c2}) are $\kappa^2_{n, \nu} = \omega^2_{n, \nu} / c^2$ where $\omega_{n, \nu}$ denotes the angular frequencies of the vibrating-membrane normal modes and $c$, the velocity of transverse vibrations propagating in the membrane. In velocity units such that $c = 1$, the eigenvalues of Eq.\ (\ref{c2}) are the squares of the angular frequencies of the vibrating-membrane normal modes.

The separation of variables gives two types of solutions. The solutions of the first type are
\begin{equation}
\eta^+_{n, \nu} (\rho, \theta) = J_\nu \left( \kappa_{n, \nu} \rho \right) \cos \nu \theta.
\label{c3}
\end {equation}
They are symmetric with respect to the reflection transformation $\theta \rightarrow - \, \theta$. Those of the second type are antisymmetric with respect to the same transformation. They are written as
\begin{equation}
\eta^-_{n, \nu} (\rho, \theta) = J_\nu \left( \kappa_{n, \nu} \rho \right) \sin \nu \theta.
\label{c4}
\end {equation}
The condition that the solution be single valued implies that $\nu = 0$, 1, 2, \dots. However, as shown by Eq.\ (\ref{c4}), the antisymmetric solutions with $\nu = 0$ do not exist. The similarity with the case of tori explains that these solutions do not exist in tori as well. Recall that in the present note, we restrict ourselves to uniform Dirichlet conditions. As in the case of tori, we use units such that the radius of the circular section be equal to 1. Then, the Dirichlet boundary conditions, which require that the solutions vanish on the circle, give $\kappa^2_{n, \nu} = x^2_{\nu,n}$, where $x_{\nu,n}$ denotes the nth zero of the Bessel function $J_\nu$, as allowed eigenvalues of the Helmholtz equation. Those of the normal modes with $n = 1$, 2, or 3 and $n = 1$ or 2 are given on the last line of Table~\ref{eigvtble}. Contrary to the case of tori, the symmetric and antisymmetric solutions with same $n$ and $\nu$ have the same eigenvalues with, of course, the exception of $\nu = 0$ for which the antisymmetric solutions do not exist.

\begin{figure}[b]
\includegraphics[trim={2.5cm 11.2cm 2cm 11.2cm},scale=0.97]{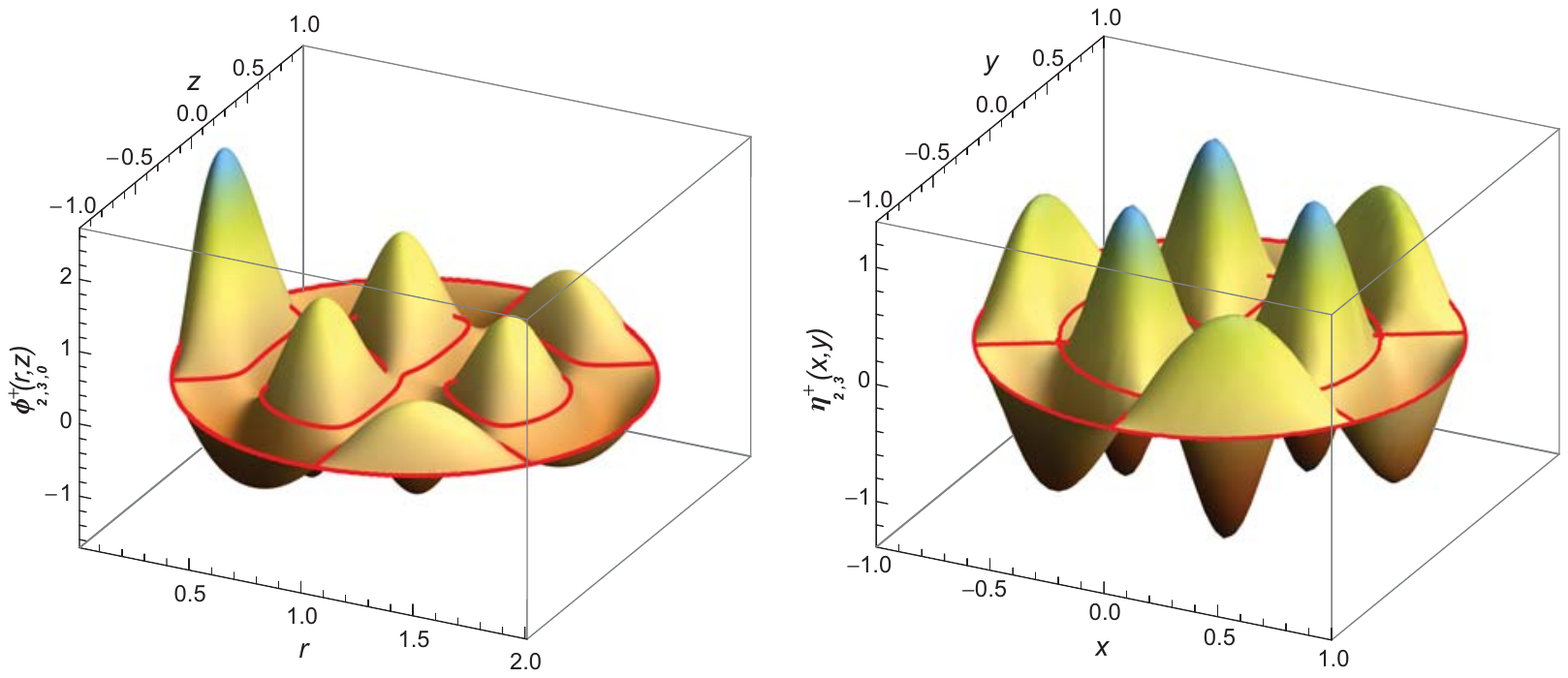}
\renewcommand{\baselinestretch}{1}
\caption{Comparison between the solutions in a torus and a cylinder. Left: torus symmetric eigenfunction with $n = 2$, $\nu = 3$, $m = 0$, and $\alpha = 1.01$; right: long-cylinder symmetric eigenfunction with $n = 2$ and $\nu = 3$. In the case of the cylinder, the coordinates are $x = \rho \cos \theta$ and $y = \rho \sin \theta$. The red curves are the contour lines $\phi = 0$ or $\eta = 0$. As before, the functions squared are normalized to 1 on the circular sections. The similarity in the number and position of the antinodes in the two graphs is apparent.}
\label{cylcomp}
\end{figure}
\normalsize
There does not seem to be an agreement on the notation to use for the indices of the eigenvalues and eigenfunctions of the Helmholtz equation in cylinders. The first index gives the number of radial antinodes and, therefore, plays a role similar to that of the principal quantum number in atomic physics. For this reason we call it $n$, as we did in the torus case. The second index specifies the symmetry of the eigenfunction with respect to rotations around the cylinder axis, so that the notation $m$ would be appropriate. However, our purpose here is the comparison with the case of tori which have a different axis of rotation and for which we have accordingly used the notation $m$ earlier. This leads us to somewhat arbitrarily choose the same notation $\nu$ for cylinders as for tori.

\begin{table}[b]
\renewcommand{\baselinestretch}{1}
\caption{Comparison of Helmholtz-equation eigenvalues in a torus and in an infinitely long right cylinder. In this latter case, the eigenvalues are the squared eigenfrequencies of a vibrating elastic circular membrane. The notations $\alpha$ and $m$ are defined in Sec.~\ref{int}.}
\begin{ruledtabular}
\begin{tabular}{ccccccc}
&$n = 1$, $\nu = 0$ & $n = 1$, $\nu = 1$ & $n = 1$, $\nu = 2$ & $n = 2$, $\nu = 0$ & $n = 2$, $\nu = 1$ & $n = 2$, $\nu = 2$\\
\hline
\multicolumn{7}{c}{$\alpha = 3$, $m = 1$}\\
\hline
$k^{+2}_{n,\nu,1}$ & 5.8697 & 14.773 & 26.466 & 30.561 & 49.315 & 70.950\\
$k^{-2}_{n,\nu,1}$ & - & 14.768& 26.465& - & 49.309& 70.950\\
\hline
\multicolumn{7}{c}{$\alpha = 3$, $m = 0$}\\
\hline
$k^{+2}_{n,\nu,0}$ & 5.7544 & 14.652 & 26.346 & 30.444 & 49.194 & 70.831\\
$k^{-2}_{n,\nu,0}$ & - & 14.654& 26.346& - & 49.195& 70.831\\
\hline
\multicolumn{7}{c}{$\alpha = 10$, $m = 1$}\\
\hline
$k^{+2}_{n,\nu,1}$ & 5.7908 & 14.690 & 26.384 & 30.481 & 49.231 & 70.867\\
$k^{-2}_{n,\nu,1}$ & - & 14.690& 26.384 & - & 49.231 & 70.868\\
\hline
\multicolumn{7}{c}{$\alpha = 10$, $m = 0$}\\
\hline
$k^{+2}_{n,\nu,0}$ & 5.7808 & 14.680 & 26.374 & 30.471 & 49.221 & 70.857\\
$k^{-2}_{n,\nu,0}$ & - & 14.680 & 26.374 & - & 49.221 & 70,858\\
\hline
\multicolumn{7}{c}{long cylinder}\\
\hline
$\kappa_{n,\nu}^2$ & 5.7832 & 14.682 & 26.375 & 30.471 & 49.219 & 70.850\\
\end{tabular}
\end{ruledtabular}
\normalsize
\label{eigvtble}
\end{table}
Let us compare the results for long cylinders with that of tori on an example. Consider, e.g., the symmetric solution of Eq.\ (\ref{c3}) with $\nu = 3$ and $n = 2$, namely $\eta^+_{2,3} (\rho, \theta) = J_3 (x_{3,2} \, \rho) \cos 3 \theta$. Compare it with the 11th symmetric eigenfunction for a torus with $m = 0$ and a ratio $\alpha = 1.01$. In this case, the torus wall almost touches the torus axis. The left graph of Fig.~\ref{cylcomp} shows this eigenfunction in the circle intersection of the torus with an axial plane, while the graph on the right shows the solution for the cylinder in the intersection with a plane normal to the cylinder axis. In spite of the extremely small value of the torus $\alpha$ ratio, there is a surprisingly high similarity between the two functions, at least from the quantitative point of view. This justifies that we use the same indices to identify the eigenfunctions of the Helmholtz equation in tori as well as in cylinders. We previously observed that the eigenvalues of the Helmholtz equation in tori should tend to those in long cylinders for $\alpha \rightarrow \infty$. This is illustrated in Table~\ref{eigvtble}, which gives the symmetric eigenvalues $k^{+2}_{n,\nu,m}$ and the antisymmetric ones $k^{-2}_{n,\nu,m}$ for $\nu = 0$, 1, and 2, $n = 1$ and 2, and $m = 0$ or 1, for $\alpha = 3$ and $\alpha = 10$ and compare them with the corresponding eigenvalues $\kappa^2_{n,\nu}$ for long cylinders.

%
%*****************************************************************************
\section{Conclusions}
\label{conc}
As stated in the introduction, the goal of this note is to provide the researchers interested in the Helmholtz equation in toroidal systems with a description of its solutions and an information on their basic properties. This should enable them to use the computer programs available on the associated web page, Ref.\ \onlinecite{hlmwp}, to obtain eigenvalues and eigenfunctions of the Helmholtz equation in accordance with their specific needs. Another goal is to provide teachers in charge of a course of lectures in mathematical physics or in numerical methods of solving partial differential equations with examples and illustrations. Our expectation is that we have reached these goals.
%
%
%
%**********************************************************************************************************
%
%
\bibliography{torus}
\end{document}